# WEB CONFERENCING TRAFFIC – AN ANALYSIS USING DIMDIM AS EXAMPLE

Patrick Seeling

Department of Computing and New Media Technologies, University of Wisconsin-Stevens Point, Stevens Point, WI 54481, USA
pseeling@uwsp.edu

## ABSTRACT

*In this paper, we present an evaluation of the Ethernet traffic for host and attendees of the popular open-source web conferencing system DimDim. While traditional Internet-centric approaches such as the MBONE have been used over the past decades, current trends for web-based conference systems make exclusive use of application-layer multicast. To allow for network dimensioning and QoS provisioning, an understanding of the underlying traffic characteristics is required.*

*We find in our exemplary evaluations that the host of a web conference session produces a large amount of Ethernet traffic, largely due to the required control of the conference session, that is heavily-tailed distributed and exhibits additionally long-range dependence. For different groups of activities within a web conference session, we find distinctive characteristics of the generated traffic.*

## 1. INTRODUCTION

Using different offerings on the World Wide Web to conduct conferences with remotely connected participants has become very popular in recent years. Used for business purposes and oftentimes in distance education settings, web conferencing enables participants to communicate, present or share screens, to name but a few of the options commonly found in web conferencing solutions available today. These different current offerings can be typically categorized into two different categories, namely (*i*) hosted by external service providers (e.g., TokBox, Adobe Acrobat Connect, DimDim) and (*ii*) self-hosted (see, e.g., OpenMeetings, BigBlueButton, DimDim). Utilizing external solutions has attracted recent attention together with the general movement of services into the 'cloud'.

Here we provide a first evaluation of the traffic characteristics for web conferencing using the open-source version of the popular DimDim conference server [1] as example. Today's web conferencing solutions make extensive use of web browsers and commonly installed plug-ins, such as Adobe Flash and Java, to allow for platform independence. This approach results in the typical operation mode of these conference systems on the application layer and they rely on a central conferencing application server that handles the related data trans-coding and forwarding tasks. This is in contrast to the traditional Internet-centric approaches, which utilize multiple Internet protocol layers in order to maximize utilization of resources on each layer. While performance evaluations for video conferencing systems in a variety of different scenarios exist, to date there is little research into the traffic characteristics of web conferencing systems, which we address with this first analysis.

This paper is organized as follows. In the remainder of this section, we review related works. In Section 2, we introduce the general operations of the DimDim open source web conference server. In Section 3, we present commonly used metrics for traffic analysis, before we present evaluation findings for exemplary web conference sessions in Section 4. In the following

Section 5, we evaluate individual activities that comprise a web conference session before we conclude in Section 6.

## 1.1 Related Works

Probably the most well-known Internet-centric approach to conferencing and collaboration is the MBONE, see, e.g., [2], whereby the conferencing here is typically enabled by individual applications. The MBONE is not web-centric, but rather relies on the interplay of members of the protocol families of RTP [3], [4] and SIP [5]–[7] in conjunction with IP multicast [8]. In contrast, browser-centric protocols are typically used for today's web conferencing systems as the latter work directly within the browser environment. These browser-based conference systems typically rely solely on TCP as transport layer protocol and HTTP in addition to other, specialized protocols on the application layer, such as RTMP and others [9], [10].

In the past, evaluations for HTTP server traffic evaluated the typical user behaviour when performing browsing tasks [11]. These evaluations date back significantly and do not necessarily reflect the amount and type of traffic that today's plug-in enhanced browsers produce. On the other hand of the spectrum is the streaming of media only, which typically has challenging demands on networks, see, e.g., [12], [13] for evaluations of currently popular video streaming applications and associated challenges.

The increased attractiveness of web-based conferencing and collaboration solutions in the recent past has started to attract research interests. The primary focus of current research efforts lies on the design and implementation of systems, see, e.g., [14] for a recent example. Evaluations that focus on the resulting network traffic are still sparse and focus on the audio and video components of these systems, see, e.g., [15], [16].An individual study evaluates a custom designed system called A-View in terms of overall bandwidth requirements, but does not provide detailed traffic statistics [17]. Prior implementations performed similar evaluations, see, e.g., [18]. Vendors of particular solutions also commonly offer the overall bandwidth requirements of their products, though again without a more detailed view of the produced traffic.

## 2. DIM DIM CONFERENCE SERVER OPERATIONS

The open-source DimDim web conference server offers similar features than those provided by commercial versions, but is based on open source streaming and media components, most predominantly the Red5 streaming media server, which offers an alternative to the Flash Media Streaming server solution offered by Adobe [9]. We note that we evaluate the traffic characteristics specifically for a web conference server which relies on a Flash Media Server alternative and that other implementations utilizing the Flash Media Server or its alternatives might operate similarly in terms of the traffic produced. Being web-based, users connect through their browser's Flash plug-in to allow for the communication between individual conference members and the server. The overall logical architecture of a conference session is that of a star topology, as the conference server acts as central connection point of all participants. We illustrate this typical topology in Figure 1. As individual participants and the host communicate with the central server, they solely utilize TCP on the transport layer. Application-layer communication is handled through different means, depending on the activities that take place. Specifically, for general configuration, file upload activities and media streaming, proprietary protocol connections on the application layer are used.

In our example case, the Real Time Messaging Protocol (RTMP) is used between an instance of the Flash player inside a user's browser and a server. RTMP is a proprietary application-layer protocol which was developed by Adobe Systems and recently made publicly available [19]. Note that RTMP itself runs on top of TCP, unlike traditional real-time oriented protocols, which commonly utilize UDP. For the hosting participant of the web conference, we also note additional frequent use of the HTTP protocol to trigger server-based events.

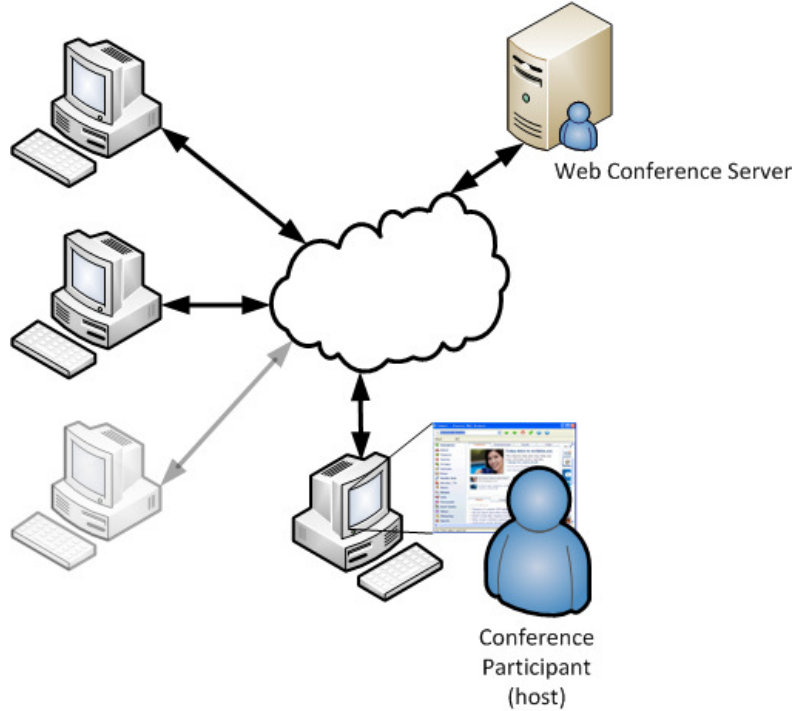

Figure 1. Generic configuration scenario for a web conference. Participants of a conference session connect with one another through a web conference application server forming a logical star topology. The conference server itself can be situated in internal or external networks.

## 3. EVALUATED TRAFFIC METRICS

For an evaluation of the traffic that occurs from participating in popular web conferencing sessions, such as those hosted or enabled by DimDim used as example throughout this paper, we performed several experiments capturing the traffic of individual participants of web-based conference sessions. The capturing of the individual traffic was performed using the popular Wireshark tool [20]. In turn, the individual sizes reported are for the complete Ethernet frames, whereby the Ethernet overhead per captured frame is 14 bytes. In the following, we introduce different statistical properties which were evaluated for the link-layer traffic of the individual participants of a web conference.

We denote the average size of the encountered $N$ Ethernet frames as $\overline{X}$ [byte] and $X_{max}$ [byte] as the maximum size of all the Ethernet frames under consideration. Subsequently, we denote the Peak-to-Mean ratio (as one measure for the variability of the underlying traffic) of the Ethernet frame sizes is defined as $PtM_X$. We calculate the variance $S_X^2$ and standard deviation $S_X$ of the Ethernet frame sizes to calculate the coefficient of variation, commonly used to measure the variability of an underlying variable, and denote it as $CoV_X$.

The Hurst parameter $H$, is one key measure of self-similarity for an underlying process and commonly used to estimate the long range dependence of a stochastic process [21], [22]. A Hurst parameter of $H$=0.5 indicates absence of self-similarity whereas $H$=1 long-range dependence. We additionally note that while the Hurst parameter is mathematically well-defined, its determination is not clearly defined, but commonly performed through estimations, see, e.g., [23]. We employ the R/S statistic [24]–[26] to estimate the Hurst parameter for the captured traffic. Specifically, we estimate the Hurst parameter $H$ as the slope of a log–log plot (also referred to as pox plot or adjusted scaled range plot) of the R/S statistic using a least squares fit. We also utilize the autocorrelation function (ACF) to evaluate the self-similarity of the network traffic by graphic means.

To illustrate the statistical behavior of the Ethernet frame sizes, we utilize the histogram (to approximate underlying probability density functions) of the Ethernet frame sizes. As the histogram for sampled values is sensitive to the bin size utilized, we use half the bin size determined by Scott in [27].

## 4. OVERALL WEB CONFERENCE TRAFFIC CHARACTERISTICS

In this section, we evaluate the traffic characteristics for three different conference sessions using the open-source DimDim web conferencing server. We provide the Ethernet frame statistics for the captured traffic in Table 1 for all of the produced Ethernet traffic for three exemplary web conferencing sessions between the host and two or three attendees. We note that in addition the minimum and maximum sizes of the captured Ethernet frames correspond to a single TCP acknowledgement and a maximum Ethernet frame load of 1500 bytes, respectively. From the presented values, we observe that the host produces or consumes approximately twice the amount of Ethernet frames than a conference attendee. Furthermore, taking the average frame sizes into account as well, we note that the host's average frame size is significantly smaller than that observed for the individual attendees. This can be explained by the required content and control packet sizes, whereby typically control packets are smaller in size and more frequently produced by the host. The same reasoning applies when comparing the traffic variability for host and attendees indicated by the peak-to-mean and coefficient of variation values in Table 1.

Table 1. Overview of Ethernet traffic produced during example web conference sessions for conference host and attendees.

| Participant | Frames $N$ | $\overline{X}$ [Byte] | $PtM_X$ | $CoV_X$ | $H$ (from R/S) |
|---|---|---|---|---|---|
| **Host** | 90530 | 248.52 | 6.092 | 1.643 | 0.872 |
| **Attendee 1** | 49449 | 336.19 | 4.503 | 1.476 | 0.821 |
| **Attendee 2** | 49449 | 336.19 | 4.503 | 1.476 | 0.821 |
| **Attendee 3** | 49257 | 332.05 | 4.56 | 1.485 | 0.836 |
| **Host** | 102402 | 132.81 | 11.4 | 1.644 | 0.826 |
| **Attendee 1** | 48534 | 166.46 | 9.095 | 1.719 | 0.757 |
| **Attendee 2** | 48822 | 167.61 | 9.033 | 1.725 | 0.769 |
| **Host** | 116998 | 231.69 | 6.535 | 1.733 | 0.812 |
| **Attendee 1** | 60370 | 325.82 | 4.647 | 1.513 | 0.821 |
| **Attendee 2** | 60544 | 324.69 | 4.663 | 1.517 | 0.827 |

The peak-to-mean ratio is significantly larger for the host, as the average frame size is smaller due to the increased number of control communication required. For the coefficient of variation ($CoV_X$), the same reasoning applies, though the differences in the variability are less pronounced. Finally, we also provide the Hurst parameter estimated from the R/S statistics pox plot in Table 1. The estimated parameter for all three conference sessions is well above 0.5, indicating the presence of long range dependency in the captured Ethernet traffic [24]–[26].

In Figure 2, we illustrate the probability density function (histogram) of Ethernet frame sizes exchanged between the host computer and the web conferencing server for all three different example web conferencing sessions. We observe that the majority of frames are fairly small, but the distribution is heavy-tailed with a second peak at the high end of maximum Ethernet frame sizes.

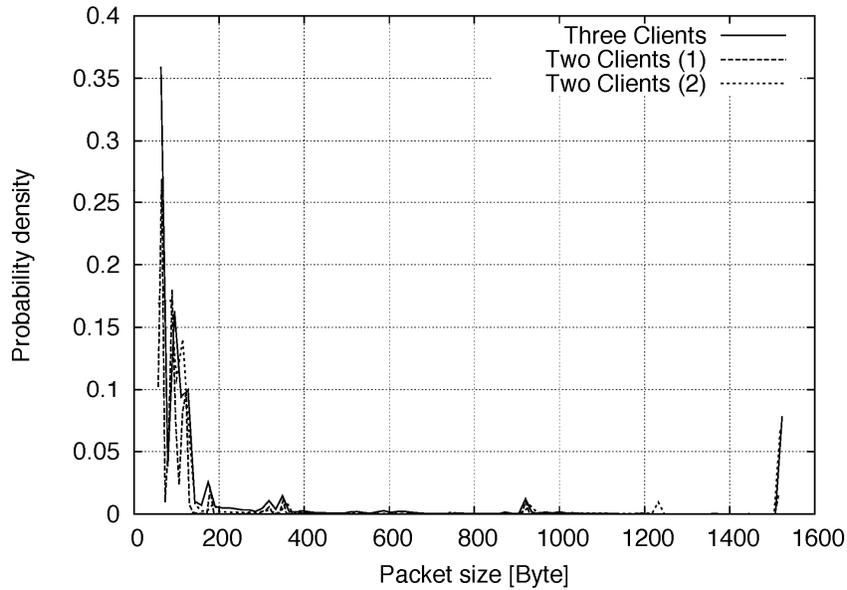

Figure 2. Probability density function (PDF) for Ethernet frame sizes in byte for the hosting participant of three exemplary web conference sessions

We illustrate the probability density function for the Ethernet frame sizes exchanged between an individual client computer and the web conference server in Figure 3.

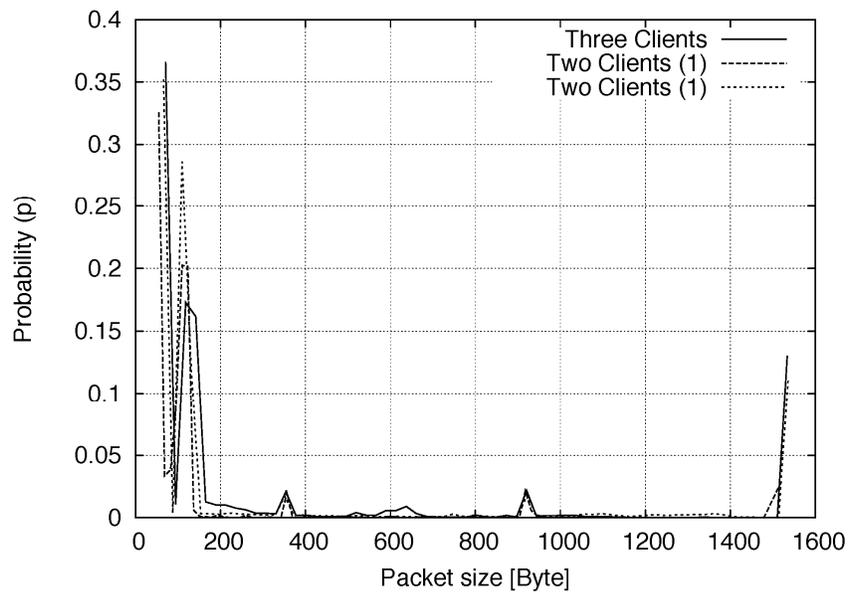

Figure 3. Probability density function (PDF) for Ethernet frame sizes in byte for one client participant of three exemplary web conference sessions.

The different distributions of Ethernet frame sizes all exhibit two peaks in the lower range of frame sizes, one for the highest frame size and are each heavy-tailed. The initial peak is around a frame size of approximately 100 bytes and the second peak is in the range of approximately 150 bytes. The final significant peak can be observed for the maximum Ethernet frame size. We note that the distributions of Ethernet traffic captured for the other participants in each individual conferencing session is almost identical to the distribution illustrated in Figure 3 and can hence be omitted for qualitative comparisons.

Next, we illustrate the Autocorrelation Function (ACF) for the host (not the server) of the web conferencing sessions in Figure 4. We observe an immediate decay of the ACF to a fairly low level, indicating the absence of self-similarity in the evaluated Ethernet frame sizes. However, as the ACF does not ultimately reach a level of zero, a low level of self-similarity is present in the frames. This low-level similarity can be attributed to the overhead present on the different protocol layers, including the aforementioned Ethernet encapsulation, which is present in all captured packets.

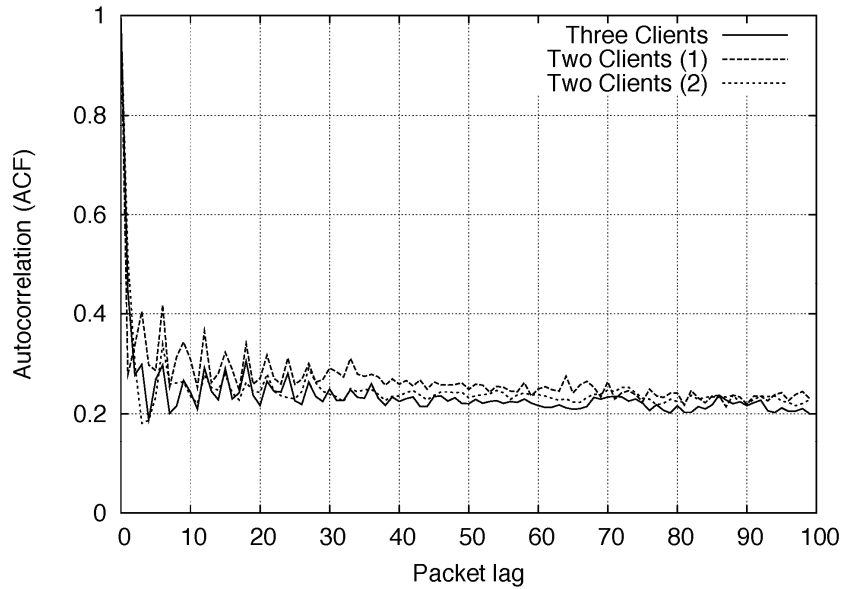

Figure 4. Autocorrelation Function (ACF) for Ethernet frame sizes for the host of three exemplary web conference sessions (all activities).

Overall, we note that the Ethernet frame traffic activity that a host participant of such conference has to manage is approximately twice the traffic activity that other attendees with low activity have to handle. This is an important characteristic that network architects need to keep in mind when dimensioning networks that are used to a large degree for supporting this particular type of traffic, e.g., distance education or corporate meetings. Furthermore, we note that the traffic we evaluated here is only for one session and for individual participants, and that the resulting traffic one has to accommodate on the server side is at least the aggregate of the Ethernet traffic captured for the participants during individual activities.

## 5. TRAFFIC CHARACTERISTICS BY ACTIVITY

In the following, we examine the different activities within the web conferencing system and their respective traffic characteristics. As overall web conferencing sessions can vary with respect to the actions performed and their durations, an ex-ante evaluation of all different activities and their respective durations is not feasible and will have to rely on large-scale studies, which are out of the scope of this paper. For the typical individual activity, however, evaluations concerning traffic characteristics can be performed and extrapolated upon similar individual activities. We present an overview of characteristics for the individual activities within an exemplary web conference session in Table 2.

Table 2. Overview of Ethernet traffic produced during example web conference sessions grouped by activities for conference host and attendees.

| Activity | Participant | Frames $N$ | $\overline{X}$ [Byte] | Fraction of host data | $PtM_X$ | $CoV_X$ | $H$ from R/S |
|---|---|---|---|---|---|---|---|
| **Audio** | Host | 8729 | 265.38 | 1 | 5.71 | 1.6 | 0.59 |
| | Attendee 1 | 7965 | 280.71 | 0.97 | 5.39 | 1.58 | 0.45 |
| | Attendee 2 | 7965 | 280.71 | 0.97 | 5.39 | 1.58 | 0.45 |
| | Attendee 3 | 7977 | 283.05 | 0.97 | 5.35 | 1.58 | 0.41 |
| **Camera** | Host | 1624 | 696.19 | 1.00 | 2.17 | 0.94 | 0.51 |
| | Attendee 1 | 1833 | 591.64 | 0.96 | 2.56 | 1.07 | 0.59 |
| | Attendee 2 | 1833 | 591.64 | 0.96 | 2.56 | 1.07 | 0.59 |
| | Attendee 3 | 1867 | 573.44 | 0.95 | 2.64 | 1.1 | 0.62 |
| **Chat** | Host | 4343 | 192.65 | 1 | 7.86 | 1.42 | 0.75 |
| | Attendee 1 | 2955 | 205.6 | 0.73 | 7.36 | 1.41 | 0.49 |
| | Attendee 2 | 2955 | 205.6 | 0.73 | 7.36 | 1.41 | 0.49 |
| | Attendee 3 | 2955 | 207.08 | 0.73 | 7.31 | 1.41 | 0.51 |
| **Presentation** | Host | 13564 | 404.26 | 1 | 3.75 | 1.41 | 0.98 |
| | Attendee 1 | 7957 | 479.6 | 0.7 | 3.16 | 1.29 | 0.85 |
| | Attendee 2 | 7957 | 479.6 | 0.7 | 3.16 | 1.29 | 0.85 |
| | Attendee 3 | 7852 | 485.88 | 0.7 | 3.12 | 1.28 | 0.9 |
| **Screen Sharing** | Host | 7674 | 367.28 | 1 | 4.12 | 1.16 | 0.79 |
| | Attendee 1 | 3206 | 685.85 | 0.78 | 2.21 | 0.95 | 0.7 |
| | Attendee 2 | 3206 | 685.85 | 0.78 | 2.21 | 0.95 | 0.7 |
| | Attendee 3 | 3248 | 690.66 | 0.8 | 2.19 | 0.94 | 0.72 |
| **Shared Browsing** | Host | 4034 | 395.13 | 1 | 3.83 | 1.23 | 0.88 |
| | Attendee 1 | 2354 | 504.89 | 0.75 | 3 | 1.07 | 0.8 |
| | Attendee 2 | 2354 | 504.89 | 0.75 | 3 | 1.07 | 0.8 |
| | Attendee 3 | 2292 | 508.79 | 0.73 | 2.98 | 1.07 | 0.78 |
| **Whiteboard** | Host | 37179 | 127.87 | 1 | 11.84 | 1.81 | 0.73 |
| | Attendee 1 | 12465 | 197.15 | 0.52 | 7.68 | 1.79 | 0.66 |
| | Attendee 2 | 12465 | 197.15 | 0.52 | 7.68 | 1.79 | 0.66 |
| | Attendee 3 | 12749 | 195.95 | 0.53 | 7.73 | 1.79 | 0.73 |

We note that the traffic characteristics differ significantly between activities. Comparing the number of Ethernet frames exchanged for different activities and their respective average sizes, we observe that different categories can be identified, namely (*i*) approximately the same number of frames and fractions of traffic as for audio and camera activities, (*ii*) slightly more frames generated by the host such as for screen sharing and shared browsing activities, (*iii*) activities such as chat and presentation where the host traffic is significantly higher than for other attendees and (*iv*) the whiteboard, where the host is the main generator and receiver of Ethernet frames.

For both audio and camera (video) activities, we note that the host traffic is approximate to the traffic observed at the other attendees. Additionally, we observe fairly high traffic variability for audio as indicated by peak-to-mean ratios and coefficients of variation, however, not for video, which is surprising given the typical behavior of encoded video. Interestingly, we also do not observe a significant level of self-similarity in the traffic, as indicated by Hurst parameter values around 0.5 or only slightly above [24]–[26].

Screen sharing and shared browsing activities are similar in the amount of aggregated traffic to the first group, but significantly different when observing the details of frame sizes and numbers generated. Here, the host exchanges significantly more frames with the web conference server (or the next hop, that is), but the average packet size is significantly lower for the host than for the attendees of the activity. For these activities, the host exchanges a large amount of additional control information with the server in addition to screen captures, resulting in the lower average frame size. The attendees, on the other hand, require only minimal control information, as they automatically will receive the screen updates from the central server, resulting in fewer frames with a larger average. We also note that while the overall traffic variability is lower than for the previous category, the Hurst parameter obtained from the R/S pox plot is higher, indicating long-range dependence in the observed traffic [24]–[26]. The long-range dependence can be explained through the nature of the activity under consideration, whereby consecutively the updates of screen content have to be sent to the central server, requiring approximately similar sized amounts of data.

Chat and presentation are similarly uni-directional for this example. For the chat activities, mainly the host was active, which additionally explains the same number of frames being captured for the other attendees. As chat messages are fairly small in size, the overall variability is fairly high as well, since the control information triggering updates at the clients is also small in size. Similarly, this can also be noted as the cause for the higher value of the Hurst parameter, indicating a low-level of long-range dependence, but only for the host. The presentation activity stands out due to the large difference of captured Ethernet frames at host and attendees. The large difference is due to (*i*) an initial upload of the document to the web conference server, where it is converted, and (*ii*) the host steering the presentation solely, resulting in additional control information that needs to be exchanged between the host and the server. Without these two characteristics, the observed frame characteristics would be close to the previous categories. Additionally, we note that the presentation activity results in the highest Hurst parameter values we obtained from the R/S statistic pox plot, strongly indicating the presence of long-range dependence in the Ethernet frame traffic observed [24]–[26]. Similar to the prior case for screen sharing, the reason for the long-range dependency can be seen in the nature of the specific activity, whereby the host switches between consecutive slides in a presentation. As this requires similar amounts of data to be exchanged with the central server, the traffic will be similar over the duration of the activity.

Finally, we note the even higher discrepancy between the Ethernet frames captured from the host versus the frames captured from attendees in the case of the whiteboard activity. We note that the drawings on the host computer are forwarded to the conference server as a combination of drawing objects and real-time activities and makes heaviest usage of the RTMP protocol [19]. In this case, the conference server aggregates some of the individual actions performed at the host before updating the clients, which results in the attendees' clients exchanging only a small fraction of RTMP packets with the server when compared to the host.

Generalized characteristics for ex-ante undetermined web conference session activities cannot be determined, as it cannot be assumed that web conferences follow the same overall patterns. For common individual activities that comprise web conferencing sessions, however, we found distinctive traffic characteristics that furthermore allow classification of different activities into distinctive groups. The grouping and group-wise traffic characteristics are useful for network dimensioning and future modelling efforts.

## 6. CONCLUSIONS

In this paper, we provided an overview of the traffic characteristics for web conferencing sessions utilizing the open-source DimDim server solution. This particular web conference server, like several others, relies on open-source implementations of Adobe's Flash ecosystem components for implementation. We find that the overall Ethernet frame traffic activity that a host participant of such conference has to manage is approximately twice the traffic activity that other attendees with low activity have to handle. We also outlined the overall traffic characteristics, which (*i*) exhibit a medium level of variability, (*ii*) follow a heavy-tailed distribution and (*iii*) exhibit a medium level of long-range dependency.

As the individual activities in web conferences are not known ex-ante, we provided an alternate view by breaking the overall conference session into individual activities, such as chat or screen sharing. We found that the individual activities can be classified into different groups, which will allow future traffic modelling activities.

In the future, we will evaluate other web conference systems with respect to the characteristics of produced traffic in a comparative manner and investigate traffic modelling for web conference systems. We will also evaluate aggregated institutional traffic for a distance education setting.

## ACKNOWLEDGEMENT

The author would like to thank the University of Wisconsin-Stevens Point University Personnel Development Committee for their publication support.

**Authors**

Patrick Seeling is an Assistant Professor in the Department of Computing and New Media Technologies at the University of Wisconsin-Stevens Point, Stevens Point, Wisconsin. He received his Dipl.-Ing. Degree in Industrial Engineering and Management from the Technical University of Berlin, Germany, in 2002. He received his Ph.D. in Electrical Engineering from Arizona State University, Tempe, Arizona, in 2005. He was a Faculty Research Associate and Associated Faculty with the Department of Electrical Engineering at Arizona State University from 2005 to 2007. Patrick Seeling published several journal articles and conference papers, as well as books, book chapters, and tutorials in the areas of multimedia, optical, and wireless networking and engineering education. He served as reviewer and technical program committee member for several journals and conferences.

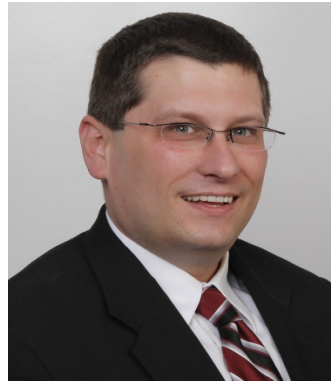